\documentclass[11pt,a4paper]{article}
\usepackage{jcappub}
\usepackage[T1]{fontenc}
\usepackage{color}
\usepackage{graphicx}
\usepackage{bm}
\usepackage{hyperref}
\usepackage[utf8]{inputenc}
\usepackage{comment}
\usepackage{amsmath,amsfonts,amssymb}

\title{Formation of supermassive primordial black holes by Affleck-Dine mechanism}

\author[a,b]{Masahiro Kawasaki}
\author[a,b]{Kai Murai}
\affiliation[a]{ Institute for Cosmic Ray Research, The University of Tokyo, Kashiwa 277-8582, Japan}
\affiliation[b]{ Kavli Institute for the Physics and Mathematics of the
  Universe (WPI), University of Tokyo,
  Kashiwa 277-8583, Japan}

\date{\today}


\abstract{We study the supermassive black holes (SMBHs) observed in the galactic centers.
Although the origin of SMBHs has not been well understood yet, previous studies suggest that seed black holes (BHs) with masses $10^{4-5}M_\odot$ exist at a high redshift ($z \sim 10$).
We examine whether primordial black holes (PBHs) produced by inhomogeneous baryogenesis can explain those seed black holes.
The inhomogeneous baryogenesis is realized in the modified Affleck-Dine mechanism.
In this scenario, there is no stringent constraint from CMB $\mu$-distortion in contrast to the scenario where Gaussian fluctuations collapse into PBHs.
It is found that the model can account for the origin of the seed BHs of the SMBHs.
}

\begin{document}
\baselineskip 0.58cm

\begin{flushright}
    IPMU 19-0093
\end{flushright}
\maketitle


\section{Introduction}

There are many evidences that most galaxies have supermassive black holes (SMBHs) with masses $10^{6}-10^{9.5}M_\odot$ in their centers~\cite{Kormendy:1995er,Magorrian:1997hw,Richstone:1998ky}.
However, the origin of SMBHs has not been well understood yet.
It is conventionally argued that stellar black holes (BHs) with masses $\mathcal{O}(10)M_\odot$ produced by stellar collapses grow to SMBHs by accretion and mergers.
But observations of high redshift QSOs indicate that SMBHs already exist at $z>6$ [e.g.~\cite{Matsuoka:2016pho}] and hence it is in dispute whether stellar BHs have time to grow to SMBHs.

Another candidate for SMBHs are primordial black holes (PBHs) which are produced from large density fluctuations in the early universe~\cite{Hawking:1971ei,Carr:1974nx,Carr:1975qj}.
Such large density fluctuations can be produced by inflation~\cite{GarciaBellido:1996qt,Yokoyama:1995ex,Kawasaki:1997ju}.
PBHs have been attracting great interest because they can be dark matter of the universe or they are a good candidate for BHs that cause gravitational wave events discovered by LIGO~\cite{Blais:2002nd,Afshordi:2003zb,Frampton:2010sw,Bird:2016dcv,Kashlinsky:2016sdv,Sasaki:2016jop,Carr:2016drx,Kawasaki:2016pql,Clesse:2016vqa,Inomata:2016rbd,Inomata:2017vxo,Ballesteros:2017fsr,Eroshenko:2016hmn,Ando:2017veq,Ando:2018nge}.
In Refs.~\cite{Bean:2002kx,10.1093/mnras/stw1679}, it is shown that BHs exsiting at $z\sim 10$ can be seeds for SMBHs if they have masses around $10^{4}-10^{5}M_\odot$.
If PBHs are formed at the cosmic temperature $\sim 1$~MeV, they have masses $\sim 10^{4-5}M_\odot$ and account for those seed BHs for SMBHs.
In fact an inflation model producing such SMBHs was proposed in Ref.~\cite{Kawasaki:2012kn}.
However, the scenario for producing supermassive PBHs has a difficulty.
PBH formation requires large density fluctuations on small scales, which leads to the CMB spectral distortion due to the Silk damping~\cite{Hu:1994bz,Chluba:2011hw}.
In fact the $\mu$-distortion of CMB gives a stringent constraint on the amplitude of the power spectrum of the curvature perturbations, $\mathcal{P}_\zeta \lesssim 10^{-4}$ for $k\sim 1-10^{4}$~Mpc, from which PBHs with masses between $4\times 10^{13}M_\odot$ and $4\times 10^{2}M_\odot$ are excluded~\cite{Kohri:2014lza}.

However, this constraint is obtained assuming that the fluctuations are nearly Gaussian and models for PBH formation based on highly non-Gaussian fluctuations can evade it.
One example is the PBH formation using the inhomogeneous Affleck-Dine baryogenesis~\cite{Dolgov:1992pu,Dolgov:2008wu,Blinnikov:2016bxu,Dolgov:2019vlq}.
(For another example, see \cite{Nakama:2016kfq}.)
Recently, Hasegawa and one of authors~\cite{Hasegawa:2017jtk,Hasegawa:2018yuy} showed that the inhomogeneous Affleck-Dine baryogenesis is realized in the framework of the minimal supersymmetric standard model (MSSM) and it produce high baryoin bubbles (HBBs) which is regions with high baryon-to-entropy ratio $\eta_b$.
Those HBBs collapse into PBHs with masses $\gtrsim 10M_\odot$ that explain the LIGO gravitational wave events.

In this paper, we examine whether the inhomogeneous Affleck-Dine baryogenesis can account for SMBHs.
It is found that the PBHs formed in the inhomogenous Affleck-Dine mechanism can be the seed BHs of the SMBHs in the galactic centers.
By varying the model parameters, we can easily control the mass distribution of the PBHs.
The scenario of Affleck-Dine mechanism depends on the SUSY breaking scheme.
We mainly discuss the gravity-mediated SUSY breaking scenario, where the baryon asymmetry in the HBBs become the massive baryons after the QCD transition and generate the density contrast.
Finally, we make some comments on the gauge-mediated SUSY breaking scenario, where the AD field fragments into the stable Q-balls. In this scenario, the Q-balls can contribute to both the seed PBHs and the dark matter abundance.

This paper is organized as follows.
In Sec.~\ref{sec:HBB}, we review the outline of the inhomogeneous Affleck-Dine baryogenesis and evaluate the distribution of the HBBs.
In Sec.~\ref{sec:collapse}, we explain how the HBBs evolve and gravitationally collapse into the PBHs.
We show the conditions for the seed BHs of the SMBHs and estimate the abundance of the PBHs in Sec.~\ref{sec:SMBH}.
Finally, we conclude in Sec.~\ref{sec:conclusion}.

\section{HBBs from Affleck-Dine field}
\label{sec:HBB}

In this section, we briefly review the generation of inhomogeneous baryon asymmetry and high baryon bubbles (HBBs) based on~\cite{Hasegawa:2017jtk,Hasegawa:2018yuy}.

\subsection{Model Setting}

In the previous works~\cite{Hasegawa:2017jtk,Hasegawa:2018yuy}, it was shown that HBBs are produced in the modified version of the AD baryogenesis in MSSM.
In this model, we modify the conventional AD baryogenesis by making two assumptions:
\begin{enumerate}
  \renewcommand{\labelenumi}{(\roman{enumi})}
  \item During inflation, the AD field has a positive Hubble induced mass, while it has a negative one after inflation.
  \item Just after inflation, the thermal potential for the AD field overcomes the negative Hubble induced mass around the origin.
\end{enumerate}
These assumptions are easily satisfied by appropriate ( and natural ) choice of the model parameters.
Under these assumptions, the potential for the AD field $\phi = \varphi e^{i\theta}$ is written as
\begin{eqnarray}
V(\phi)=
\left\{ \begin{array}{ll}
\left( m_{\phi}^2 + c_I H^2 \right) |\phi|^2 + V_{\mathrm{NR}},
& (\mathrm{during~inflation}) \\
\left( m_{\phi}^2 - c_M H^2 \right) |\phi|^2 + V_{\mathrm{NR}} + V_{\mathrm{T}}(\phi),
& (\mathrm{after~inflation}) \\
\end{array} \right.
\end{eqnarray}
where $c_I, c_M$ are dimensionless positive constants, $m_{\phi}$ is the soft SUSY breaking mass for the AD field, and $V_{\mathrm{NR}}$ represents the non-renormalizable contribution given by
\begin{equation}
  V_{\mathrm{NR}} =
  \left(
  \lambda a_M \frac{m_{3/2}\phi^n}{nM_{\mathrm{Pl}}^{n-3}} + \mathrm{h.c.}
  \right) +
  \lambda^2 \frac{|\phi|^{2(n-1)}}{M_{\mathrm{Pl}}^{2(n-3)}},
\end{equation}
where $\lambda, a_M$ are dimensionless constants.
The integer $n$ ($\geq 4$) is determined by specifying the MSSM flat direction.
$V_{\mathrm{T}}$ is the thermal potential for the AD field induced by the thermalized decay product of the inflaton and is written as
\begin{eqnarray}
V_{\mathrm{T}}(\phi)=\left\{ \begin{array}{ll}
c_1T^2|\phi|^2,
& |\phi| \lesssim T, \\
c_2T^4\ln \left( \frac{|\phi|^2}{T^2} \right),
& |\phi| \gtrsim T, \\
\end{array} \right.
\end{eqnarray}
where $c_1,c_2$ are $\mathcal{O}(1)$ parameters relevant to the coupling between the AD field and the thermal bath.

\subsection{Dynamics of the AD field}

\begin{figure}[tb]
  \begin{center}
    \includegraphics[clip,width=0.5\textwidth]{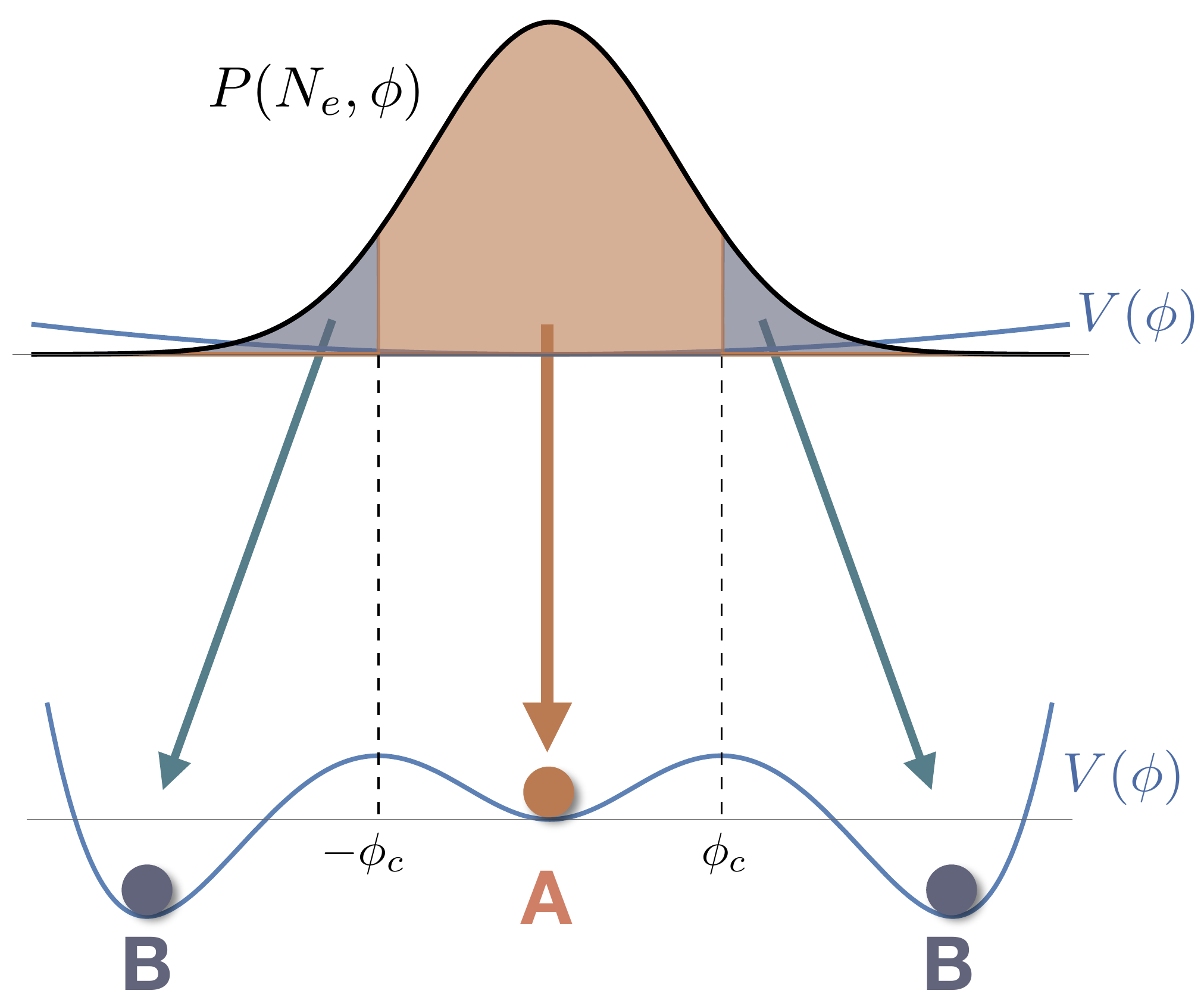}
    \caption{The schematic view of the dynamics of the AD field during and after inflation. {\it Upper side}: During inflation, the coarse-grained AD field diffuses in the complex plain and takes different values in different Hubble patches. {\it Lower side}: Just after inflation, the multi-vacuum structure appears due to the thermal potential and the negative Hubble induced mass. In the patches with $|\phi|>\phi_c$, the AD field rolls down to the vacuum B. On the other hand, the patches with $|\phi|<\phi_c$, the AD field rolls down to the vacuum A.}
    \label{fig:vacua}
  \end{center}
\end{figure}

Let us describe the dynamics of the AD field in this scenario.
During inflation, the potential of the AD field has the minimum at $\varphi = 0$ due to the positive Hubble induced mass.
Furthermore, when $c_I < 1$, the AD field acquires quantum fluctuations around it.
Therefore, the AD field coarse-grained over the Hubble scale stochastically fluctuates and takes a different value in each Hubble patch shown in Fig.~\ref{fig:vacua}.

After inflation, by the assumption (ii), the negative Hubble induced mass and the thermal potential bring on the multi-vacuum structure with the vacuum ``A'' at $\varphi=0$ and the vacuum ``B'' at $\varphi \neq 0$ (see Fig.~\ref{fig:vacua}).
Then, the AD field rolls down the potential to one of the two vacua depending on values of the AD field at the end of inflation $t_e$.
If $\varphi(t_e)$ in some patch is smaller than the maximal point between the two vacua $\varphi_c (t_e)$, the AD field rolls down to the vacuum A soon after inflation
Thus, in this case almost no baryon number is produced, i.e.
\begin{equation}
  \eta _b^{(A)} \simeq 0.
\end{equation}
On the other hand, in the patches satisfying $\varphi(t_e) > \varphi_c (t_e)$, the AD field rolls down to the vacuum B.
The vacuum B vanishes later due to the thermal potential or soft SUSY breaking mass term and the conventional AD baryogenesis takes place.
Therefore, the AD field begins to oscillate around the origin at $H(t) \simeq H_{\mathrm{osc}}$, which produces the baryon asymmetry given by
\begin{align}
  \eta_b^{(B)} &\simeq \epsilon \frac{T_{R}m_{3/2}}{H^2_{\mathrm{osc}}}
  \left(
  \frac{\varphi_{\mathrm{osc}}}{M_{\mathrm{Pl}}}
  \right)^2,\\
  \epsilon &= \sqrt{\frac{c}{n-1}}\frac{q_b |a_M| \sin (n\theta_0+\arg (a_M))}
  {3\left( \frac{n-4}{n-2}+1 \right)},
\end{align}
where the subscript ``osc'' represents the values evaluated at the onset of the oscillation, $T_R$ is the reheating temperature, $q_b$ is the baryon charge of the AD field, and $\theta_0$ is the initial phase of the AD field.
In this way, the inhomogeneous AD baryogenesis takes place and the HBBs with large baryon asymmetry are formed.

Here, we derive the condition for our assumption (ii).
When the decay products are thermalized instantaneously, the cosmic temperature is given by
\begin{equation}
    \label{eq:cosmic_temp}
  T^{\mathrm{inst}}(t) \simeq \left( T^2_RH(t)M_{\mathrm{Pl}} \right)^{1/4}.
\end{equation}
With use of Eq.~(\ref{eq:cosmic_temp}), the assumption (ii) is satisfied if
\begin{equation}
  \Delta \equiv \frac{T^2_RM_{\mathrm{Pl}}}{H(t_e)^3} \gtrsim 1,
\end{equation}
where we set $c_M, c_1 \sim 1$ for simplicity.
The critical point just after inflation $\varphi_c(t_e)$ can be written as
\begin{equation}
  \varphi_c(t_e) \equiv \varphi_c = \Delta^{1/2}H(t_e),
\end{equation}
where we also set the $\mathcal{O}(1)$ model parameters to unity.

\subsection{Volume fraction of HBBs}

Next, we analytically evaluate the volume fraction of the HBBs.
The evolution of the coarse-grained AD field during inflation is described by the Langevin equation including the Gaussian noise~\cite{Vilenkin:1982wt,Starobinsky:1982ee,Linde:1982uu} and the probability distribution function of the AD field with respect to $e$-folding number $N (\propto \ln a ~[ a\text{: scale factor}])$, $P(N,\phi)$ follows the Fokker-Planck equation:
\begin{equation}
  \frac{\partial P(N,\phi)}{\partial N} =
  \sum_{i = 1,2} \frac{\partial}{\partial \phi_i}
  \left[
  \frac{\partial V(\phi)}{\partial \phi_i}\frac{P(N,\phi)}{3H^2} +
  \frac{H^2}{8\pi^2} \frac{\partial P(N,\phi)}{\partial \phi_i}
  \right],
\end{equation}
where $(\phi_1,\phi_2) = (\Re [\phi], \Im [\phi])$.
The first term in the RHS is classically induced by the potential and the second term represents the quantum fluctuations.
Assuming the initial condition $P(0,\phi) = \delta(0)$ and the constant Hubble parameter during inflation $H(t \leq t_e) = H_I$, we obtain
\begin{align}
  P(N,\phi)
  &= \frac{1}{2\pi\sigma^2(N)} e^{-\frac{\varphi^2}{2\sigma^2(N)}},\\
  \sigma^2(N)
  &= \left(\frac{H_I}{2\pi}\right)^2 \frac{1-e^{-c_I'N}}{c_I'},
\end{align}
where we have used $V(\phi) \simeq c_IH_I^2\phi^2$ and defined $c_I' \equiv (2/3)c_I$.
The phase of the AD field $\theta$ is random unless large CP violation terms such as Hubble induced A-terms are introduced.

\begin{figure}[tb]
  \begin{center}
    \includegraphics[clip,width=0.48\textwidth]{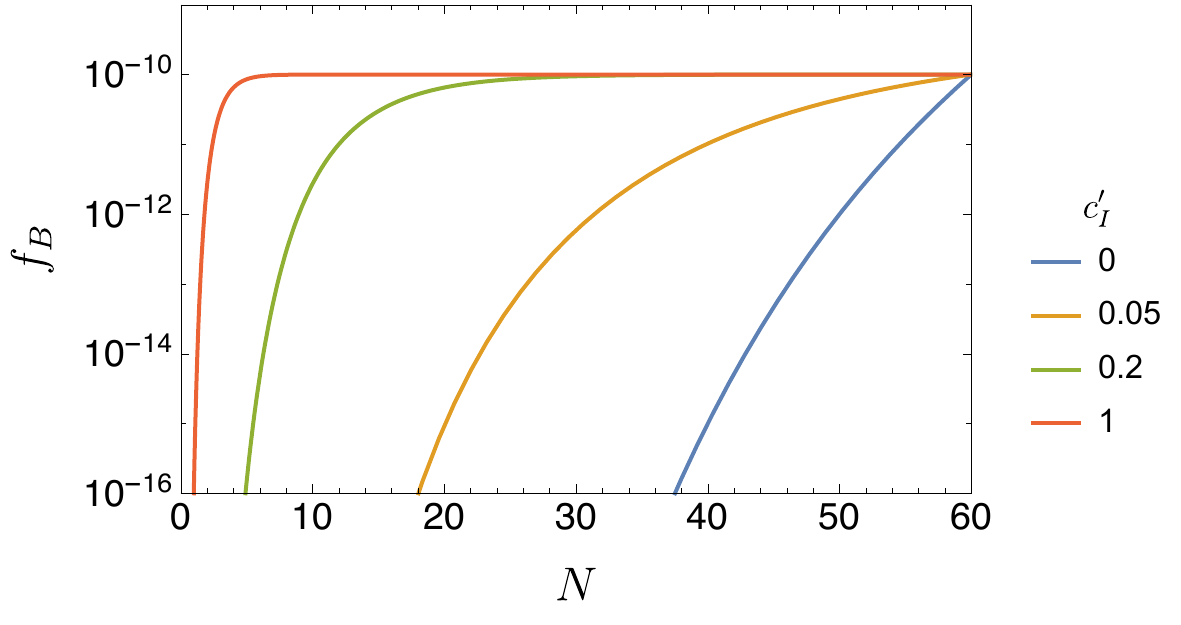}
    \includegraphics[clip,width=0.48\textwidth]{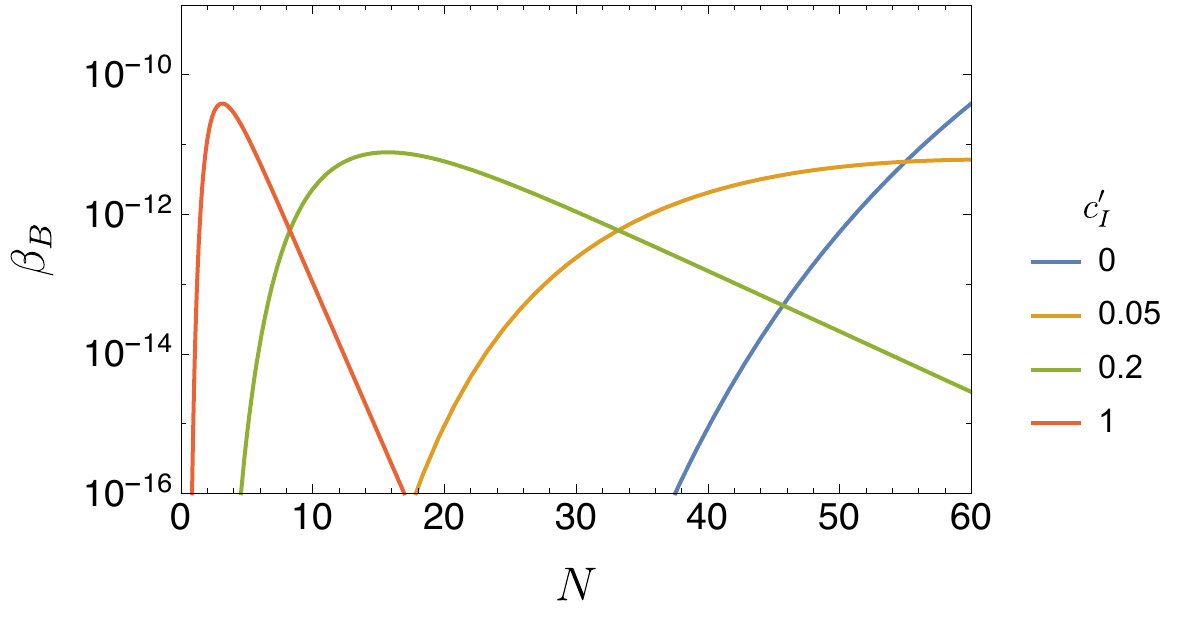}
    \caption{The evolutions of the volume fraction of HBBs $f_B(N)$ (left) and the production rate of HBBs at $\beta_B(N)$ (right) for various $c_I'$.
    $\Delta$ is adjusted for each plot so that $f_B(N=60)$ takes the same value.}
    \label{fig:fbeta}
  \end{center}
\end{figure}

As discussed above, the patches where the AD field rolls down to the vacuum B cause the AD baryogenesis and form HBBs.
Therefore, at the $e$-folding number $N$, the physical volume of the regions which would later become HBBs, $V_B(N)$ is evaluated as
\begin{equation}
  V_B(N) = V(N) \int_{\varphi>\varphi_c}P(N,\phi)\mathrm{d}\phi \equiv V(N) f_B(N).
\end{equation}
Here, we represent the physical volume of the Universe at $N$ as $V(N) \sim r_H^3e^{3N}$, where $r_H \sim H_I^{-1}$ is the Hubble radius during inflation.
The volume fraction of the regions which would later become HBBs, $f_B(N)$, is evaluated as
\begin{equation}
  f_B(N) = \int_0^{2\pi} \mathrm{d}\theta \int_{\varphi_c}^\infty \mathrm{d}\varphi \varphi \frac{e^{-\frac{\varphi^2}{2\sigma^2(N)}}}{2\pi\sigma^2(N)}
  = e^{-\frac{2\pi^2\Delta}{\tilde{\sigma}^2(N)}},
\end{equation}
where $\tilde{\sigma}^2(N) \equiv \left(1-e^{-c_I'N}\right)/c_I'$.
We show the evolutions of $f_B(N)$ in Fig.~\ref{fig:fbeta} (left panel).

\subsection{Size distribution of HBBs}

The creation rate of the regions which would later become HBBs is obtained by differentiating $V_B(N)$ with respect to $N$ as
\begin{equation}
  \frac{\mathrm{d}V_B(N)}{\mathrm{d}N} = 3V_B + V(N) \int_{\varphi > \varphi_c}\frac{P(N,\phi)}{\mathrm{d}N}\mathrm{d}\phi.
\end{equation}
The first term in the RHS represents the growth of the existing regions which would later become HBBs due to the cosmic expansion and the second term represents the creation of such regions.
Therefore, the fraction of the HBBs with the scale exiting the horizon at $N$ is evaluated at the end of inflation $N_e$ as
\begin{align}
  \beta_B(N;N_e) &= \frac{1}{V(N_e)}\cdot e^{3(N-N_e)}
  \left[
  V(N) \int_{\varphi>\varphi_c}
  \frac{\mathrm{d}P(N,\phi)}{\mathrm{d}N} \mathrm{d}\phi
  \right] \nonumber\\
  &= \frac{\mathrm{d}}{\mathrm{d}N}f_B(N) =
  \frac{\left( \pi c_I' \right)^2 \Delta}{\sinh \left( c_I'N/2 \right)}f_B(N).
\end{align}
We show the evolutions of $\beta_B(N)$ in Fig.~\ref{fig:fbeta} (right panel).

The scale of the HBBs $k$ can be expressed in terms of the $e$-folding number $N$ when the scale $k$ exits the horizon as
\begin{equation}
  k(N) = k_*e^{N-N_{\mathrm{CMB}}},
\end{equation}
where $k_*$ is the CMB pivot scale and $N_{\mathrm{CMB}}$ is the $e$-folding number when the pivot scale exits the horizon.
The horizon mass when the scale $k$ re-enter the horizon is evaluated as
\begin{equation}
  M_H \simeq 19.3M_{\odot}
  \left( \frac{g_*}{10.75} \right)^{-1/6}
  \left( \frac{k}{10^6\mathrm{Mpc}^{-1}} \right)^{-2},
\end{equation}
where $M_{\odot}$ is the solar mass and $g_*$ is the effective degree of freedom of relativistic particles.
Equivalently, the $e$-folding number $N$ is evaluated in terms of $M_H$ as
\begin{equation}
  N(M_H) \simeq -\frac{1}{2}\ln \frac{M_H}{M_{\odot}} + 21.5 + N_{\mathrm{CMB}},
  \label{eq:NMassRelation}
\end{equation}
where we used $g_* = 10.75$ and $k_* = 0.002 ~\mathrm{Mpc}^{-1}$.
It is known that the typical inflationary models suggest that
\begin{equation}
  N_e - N_{\mathrm{CMB}} \sim 50 - 60.
\end{equation}
On the other hand, to solve the horizon and flatness problem of the Big Bang cosmology, the total number of $e$-foldings of the inflation era should be
\begin{equation}
  N_e \gtrsim 60.
\end{equation}
In the rest of this paper, we fix $N_e = 60$ and $N_{\mathrm{CMB}} = 10$.
It is also convenient to relate the cosmic temperature $T$ to the horizon mass as
\begin{equation}
  T(M_H) = 434\mathrm{MeV}\left( \frac{M_H}{M_{\odot}} \right)^{-1/2}.
  \label{eq:TempMassRelation}
\end{equation}

\section{Gravitational collapse of HBBs}
\label{sec:collapse}

In this section, we discuss the evolution of the HBBs and their gravitational collapse into PBHs.

\subsection{Density contrast of HBBs}

Just after inflation, the energy density inside and outside the HBBs are almost the same because of the energy conservation of the AD field and the domination of the oscillation energy of the inflaton.
After the AD field decays, the quarks carry the produced baryon asymmetry.
As long as the quarks remain relativistic, the density fluctuations are not generated.
After the QCD phase transition, the baryon number is carried by massive baryons (protons and neutrons).
Since the energy of the baryons behaves as non-relativistic matter, their energy density is given by $\rho \simeq n_bm_b$ ($m_b$:  the nucleon mass).
Thus, the density contrast between inside and outside the HBBs is written as
\begin{equation}
  \delta \equiv \frac{\rho^{\mathrm{in}}-\rho^{\mathrm{out}}}{\rho^{\mathrm{out}}}
  \simeq \frac{n_b^{\mathrm{in}}m_b}{\left(\pi^2/30\right)g_*T^4}
  \simeq 0.3\eta_b^{\mathrm{in}}\left( \frac{T}{200{\mathrm{MeV}}} \right)^{-1} \theta\left( T_{\mathrm{QCD}}-T \right),
  \label{eq:densitypert}
\end{equation}
where $T_\text{QCD}$ is the cosmic temperature at the QCD phase transition and $\theta(x)$ is the Heaviside theta function.
Here, we have used $m_b \simeq 938$~MeV.

Here, we make three comments.
First, HBBs are considered as top-hat type baryonic isocurvature fluctuations.
Such small-scale isocurvature fluctuations are hardly constrained by the CMB observations.\footnote{
Large isocurvature baryonic perturbations are constrained from the big-bang nucleosynthesis if they are Gaussian~\cite{Inomata:2018htm}.}
Although isocurvature fluctuations induce adiabatic ones after the QCD phase transition, the produced perturbations are highly non-Gaussian.
In addition, HBBs are presumed to be rare objects and adiabatic perturbations averaged over the whole universe are small.
Therefore, this model does not suffer from the stringent constraints from CMB $\mu$-distortion and the PTA experiments~\cite{Arzoumanian2015,Lentati2015,Shannon2015}.
Second, the baryon asymmetry $\eta_b^{\mathrm{in}}$ is different from HBB to HBB depending on the initial phase of the AD field $\theta_0$, which is efficiently random due to the flatness of the phase direction.
Although this effect does not bring a substantial change to the following discussion, we assume that all HBBs have the baryon asymmetry for simplicity.
Actually this assumption can be realized by introducing the Hubble induced A-term.
Third, we implicitly assumed that the AD field decays into the quarks above.
However, it is known that the coherent oscillation of the AD field usually fragments to the localized lumps called Q-balls~\cite{Kusenko:1997si,Enqvist:1997si,Kasuya:1999wu}.
In this paper, we mainly focus on the gravity-mediated SUSY breaking scenario, where the produced Q-balls are unstable and decay into the quarks.
On the other hand, in the gauge-mediated SUSY breaking scenario, the produced Q-balls are stable and can make density fluctuations which become PBHs later.
We will make a comment on this case later.

\subsection{PBH formation}

From Eq.(\ref{eq:densitypert}), the HBBs become over-dense after the QCD transition.
If the density contrast is large enough, the overdense regions collapse into PBHs after they re-enter the horizon.
In the radiation-dominated era, the threshold value of the density contrast for the PBH formation is roughly estimated as $\delta_c \simeq w$ with $w \equiv p/\rho$~\cite{Carr:1974nx}, which we adopt in this paper.
Notice that our model is hardly sensitive to the choice of estimation of $\delta_c$ unlike the case of the Gaussian density perturbations.

In the present case $w$ is written in terms of $\delta(T)$ as
\begin{equation}
  w(T) = \frac{p^{\mathrm{in}}}{\rho^{\mathrm{in}}}
  \simeq \frac{p^{\mathrm{out}}}{\rho^{\mathrm{in}}}
  = \frac{1}{3}\frac{1}{1+\delta(T)}.
\end{equation}
Thus, the condition for the PBH formation $\delta(T) > \delta_c(T) \simeq w(T)$ is given by
\begin{equation}
  \delta(T) \gtrsim \frac{1}{3}\frac{1}{1+\delta(T)} \Longleftrightarrow
  \delta(T) \gtrsim 0.26,
\end{equation}
and this gives the upper bound of the temperature at the horizon re-entry for the PBH formation.
The critical temperature for the PBH formation $T_c$ is obtained from Eq.(\ref{eq:densitypert}):
\begin{equation}
  T_c \simeq \mathrm{Min} \left[ 231\eta_b^{\mathrm{in}}\mathrm{MeV}, T_{\mathrm{QCD}} \right],
\end{equation}
which leads to the lower bound of the horizon mass for the PBH formation,
\begin{equation}
  M_c \simeq \mathrm{Max}
  \left[
  14.1 \left( \eta_b^{\mathrm{in}} \right)^{-2}M_{\odot} ,
  18.8M_{\odot} \left( \frac{T_{\mathrm{QCD}}}{200\mathrm{MeV}} \right)^{-2}
  \right].
\end{equation}
Here we used Eq.(\ref{eq:TempMassRelation}).
Since the mass of the formed PBH is comparable with the horizon mass at the horizon re-entry, $M_{\mathrm{PBH}} \sim M_H$, the mass distribution of the PBHs is given as
\begin{equation}
  \beta_{\mathrm{PBH}}(M_{\mathrm{PBH}}) = \frac{1}{2}\beta_B
  \left(
  N(M_{\mathrm{PBH}})
  \right)
  \theta(M_{\mathrm{PBH}}-M_c),
  \label{eq:betarelation}
\end{equation}
where $\beta_{\mathrm{PBH}}$ is the volume fraction of the HBBs which become the PBHs with a mass $M_{\mathrm{PBH}}$ at the horizon re-entry over logarithmic mass interval $\mathrm{d}\left( \ln M_{\mathrm{PBH} } \right)$.
The relation between $N$ and $M_H$, Eq.(\ref{eq:NMassRelation}) introduces the factor $1/2$ in Eq.(\ref{eq:betarelation}).

\section{SMBH seeds from the modified Affleck-Dine baryogenesis}
\label{sec:SMBH}

In this section, we investigate the possibility that the PBHs in our scenario account for the SMBHs.
First, let us summarize the conditions for the seed BHs of the SMBHs.
As supported by observations of high red-shift ($z \sim 10$) SMBHs, the formation of galaxies is considered to be preceded by the formation of seed BHs~\cite{vandenBosch:2012uv}.
Therefore, we assume that the number density of SMBHs is equal to that of galaxies.
The number density of galaxies $N_{\mathrm{gal}}$ is not well-known.
Here, we take it as
\begin{equation}
    N_{\mathrm{gal}} = (10^{-3}-10^{-1})~\mathrm{Mpc}^{-3}.
\end{equation}
This is consistent with the estimated values in Refs.~\cite{Conselice_2016,Shinkai:2016xya}.
In Ref.~\cite{10.1093/mnras/stw1679}, it is shown in the numerical simulation that PBHs with masses around $(10^4-10^5)M_{\odot}$ subsequently grow up to $10^9M_{\odot}$.
Since PBHs heavier than $(10^4-10^5)M_{\odot}$ are also supposed to become SMBHs~\cite{Dolgov:2017aec}, we assume that PBHs heavier than a certain boundary value $M_b$ are  seed BHs of  SMBHs.
Here, we take $M_b$ as
\begin{equation}
  M_b = (10^4-10^5)M_{\odot},
\end{equation}
following Ref.~\cite{Dolgov:2019vlq}.

\begin{figure}[tb]
  \begin{center}
    \includegraphics[clip,width=0.48\textwidth]{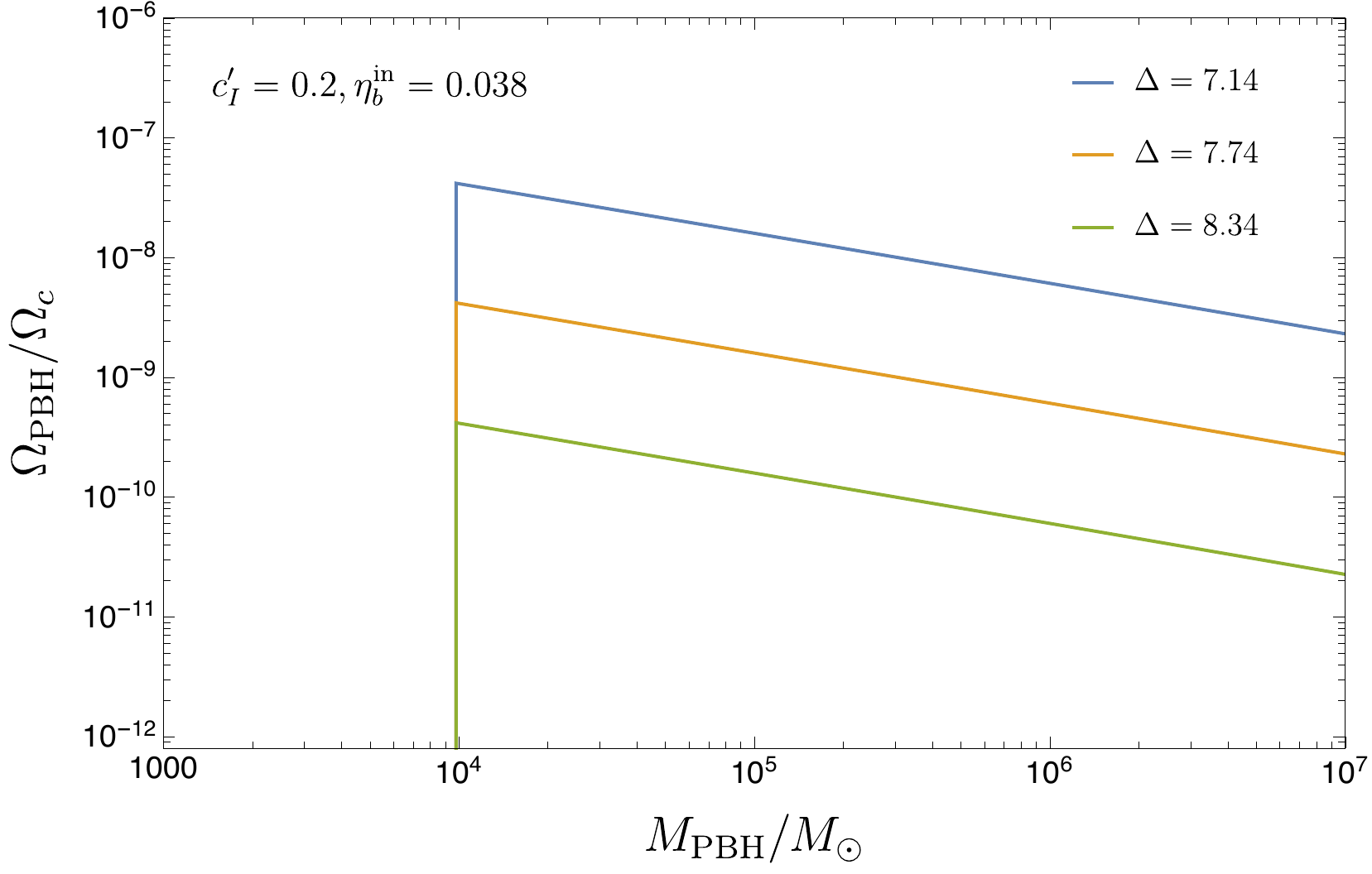}
    \includegraphics[clip,width=0.48\textwidth]{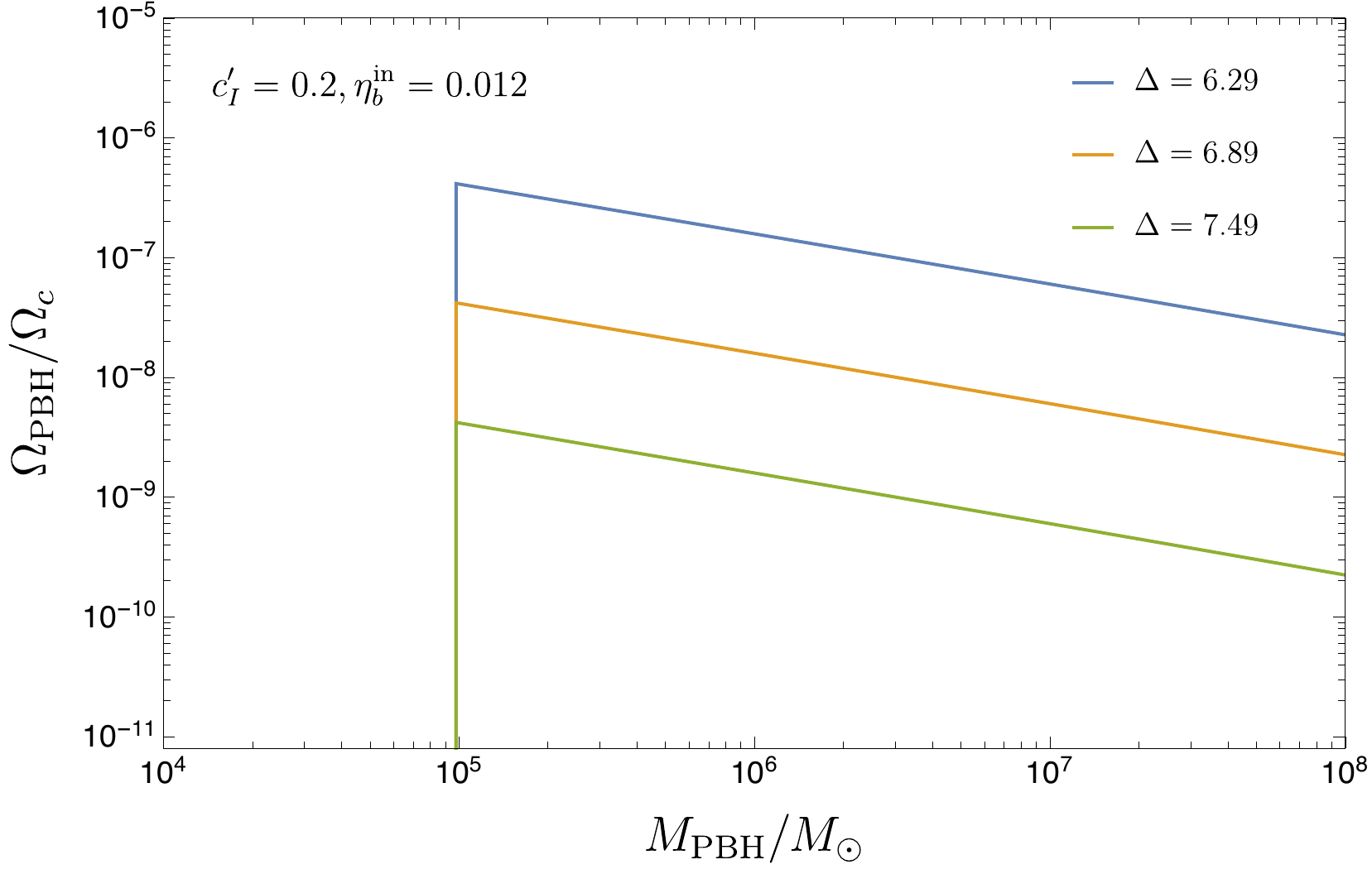}
    \caption{The PBH abundance.
            The left panel is the case with $M_c = 10^4M_{\odot}$ and the right panel is the case with $M_c = 10^5M_{\odot}$.
        In each panel, three lines correspond to $N_{\mathrm{gal}} = 0.1,0.01,0.001$~Mpc$^{-3}$ from top to bottom if $M_c \geq M_b$.}
    \label{fig:loglog}
  \end{center}
\end{figure}

Next, let us estimate the abundance of the PBHs.
The present abundance of the PBHs with mass $M_{\mathrm{PBH}}$ over logarithmic mass interval $\mathrm{d}(\ln M_{\mathrm{PBH}})$ is given by
\begin{align}
  \frac{\Omega_{\mathrm{PBH}} (M_{\mathrm{PBH}})}{\Omega_c}
  &\simeq \left. \frac{\rho_{\mathrm{PBH}}}{\rho_m}\right|_{\mathrm{eq}}
  \frac{\Omega_m}{\Omega_c}
  = \frac{\Omega_m}{\Omega_c}\frac{T(M_{\mathrm{PBH}})}{T_{\mathrm{eq}}}
  \beta_{\mathrm{PBH}}(M_{\mathrm{PBH}}) \nonumber\\
  &\simeq
  \left(
  \frac{\beta_{\mathrm{PBH}}(M_{\mathrm{PBH}})}{1.6\times 10^{-9}}
  \right)
  \left(
  \frac{\Omega_ch^2}{0.12}
  \right)^{-1}
  \left(
  \frac{M_{\mathrm{PBH}}}{M_{\odot}}
  \right)^{-1/2},
\end{align}
where $\Omega_c$ and $\Omega_m$ are the present density parameters of dark matter and matter, respectively.
$h$ is the present Hubble parameter in units of $100$~km/sec/Mpc.
$T(M_{\mathrm{PBH}})$ and $T_{\mathrm{eq}}$ are respectively the temperatures at the formation of PBHs with a mass $M_{\mathrm{PBH}}$ and at the matter-radiation equality.
The number density of PBHs over $\mathrm{d}(\ln M_{\mathrm{PBH}})$ is written as
\begin{align}
  \frac{\mathrm{d}N_{\mathrm{PBH}}}{\mathrm{d}(\ln M_{\mathrm{PBH}})}(M_\mathrm{PBH})
  &= \frac{\Omega_{\mathrm{PBH}}(M_{\mathrm{PBH}})\rho_{\mathrm{crit}}}
  {M_{\mathrm{PBH}}} \\
  &\simeq 3.3 \times 10^{10}\mathrm{Mpc}^{-3}   \left(\frac{\beta_{\mathrm{PBH}}(M_{\mathrm{PBH}})}{1.6\times 10^{-9}}\right)
  \left(
  \frac{M_{\mathrm{PBH}}}{M_{\odot}}
  \right)^{-3/2},
\end{align}
where $\rho_{\mathrm{crit}} \equiv 3H_0^2M_{\mathrm{Pl}}^2$ is the critical density.
The relation which $N_{\mathrm{gal}}$ and $M_b$ should satisfy is written as
\begin{equation}
  N_{\mathrm{gal}} = \int_{\ln M_b}^{\infty}\mathrm{d}(\ln M_{\mathrm{PBH}}) \frac{\mathrm{d}N_{\mathrm{PBH}}}{\mathrm{d}\ln (M_{\mathrm{PBH}})}.
  \label{eq:NgalCondition}
\end{equation}
In Fig.~\ref{fig:loglog}, we show the abundance of the PBHs with the parameters satisfying Eq.~(\ref{eq:NgalCondition}).
Although the number density of galaxies and the masses of the seed BHs are not well-known and we put some assumptions on them in the above discussion, the mass distribution of the PBHs can be easily modified in our model by varying the model parameters $(c_I, \Delta, \eta_b^{\mathrm{in}})$.

\begin{figure}[tb]
  \begin{center}
    \includegraphics[clip,width=0.8\textwidth]{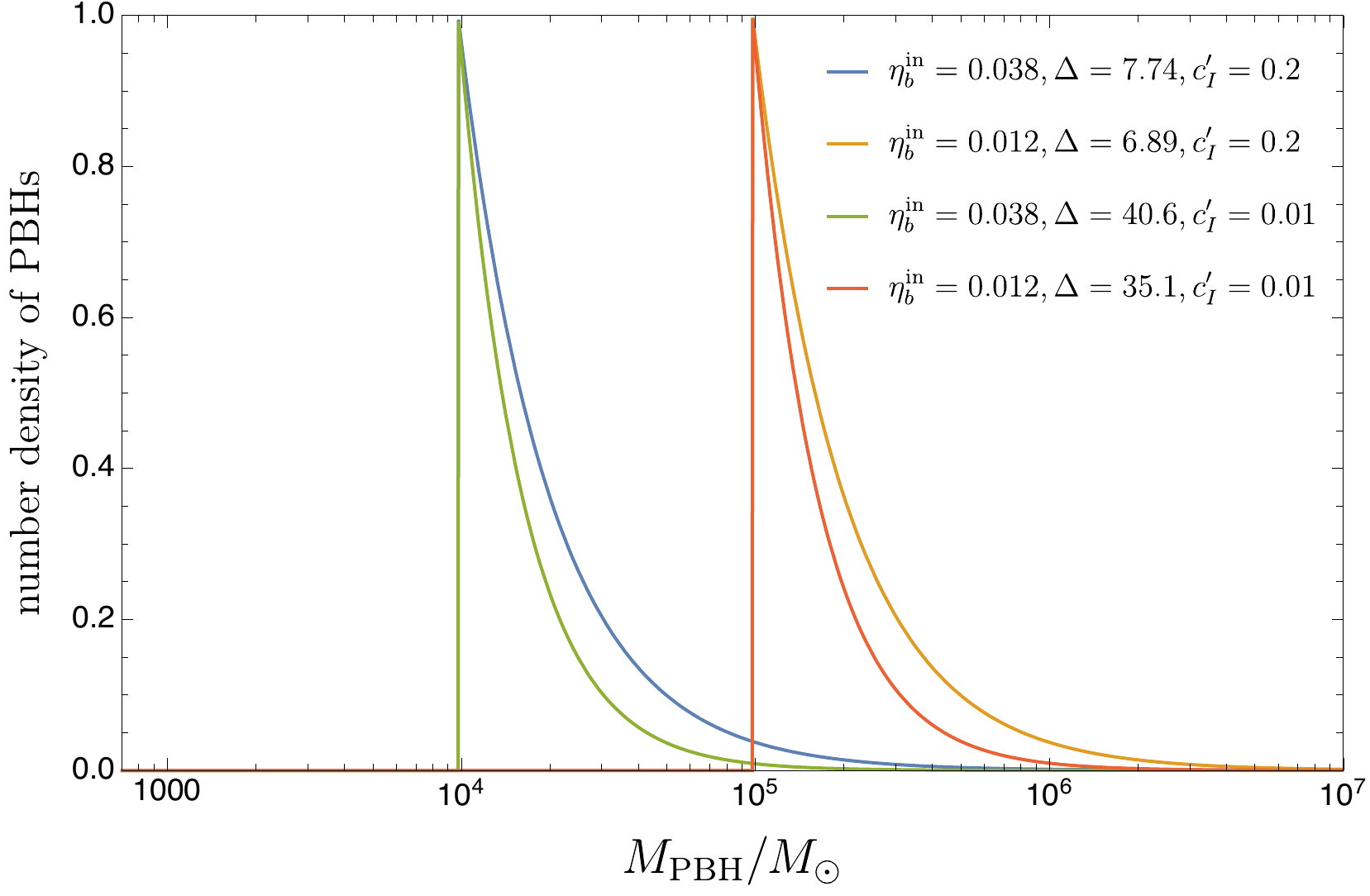}
    \caption{The number density of PBHs assuming $M_c \geq M_b$.
            Each plot is normalized by its maximum value.
            If $M_c<M_b$, the cutoff below which there is no PBH appears at $M_{\mathrm{PBH}}=M_b$.}
    \label{fig:loglin}
  \end{center}
\end{figure}

The HBBs with masses less than $M_c$ do not collapse into PBHs but they can contribute to the baryon asymmetry of the Universe.
In order to evaluate this contribution, we introduce the quantity
\begin{equation}
  \eta^B_b = \big( f_B(N_e)-f_B(N(M_c)) \big)\eta_b^{\mathrm{in}},
\end{equation}
which represents the averaged baryon asymmetry.
Those baryons are highly inhomogeneous and nucleosynthesis in the HBBs are different from the standard BBN because of high baryon density environment, which predicts quite different abundances for light elements.
Therefore, in order not to spoil the success of the standard BBN, we should require $\eta^B_b \ll \eta_b^{\mathrm{ob}}\sim 10^{-10}$ where $\eta_b^{\mathrm{ob}}$ is the observed baryon asymmetry.
The parameters in Fig.~\ref{fig:loglog}, where $\eta_b^B/\eta_b^{\mathrm{ob}} \lesssim 10^{-4}$, satisfy this requirement.
In Fig.~\ref{fig:loglin}, we show the number density spectrum of the PBHs.
It can be seen that they have similar shapes with sharp peaks at $M_c$ independent of the choice of the parameters.
Since the HBBs can account only for the small fraction of the observed baryon asymmtery, we need another baryogenesis mechanism.
The simplest possibility is to utilize  another AD field (flat direction) with negative Hubble mass during and after inflation, which leads to the conventional Affleck-Dine baryogenesis.

\begin{figure}[htbp]
  \begin{center}
    \includegraphics[clip,width=0.48\textwidth]{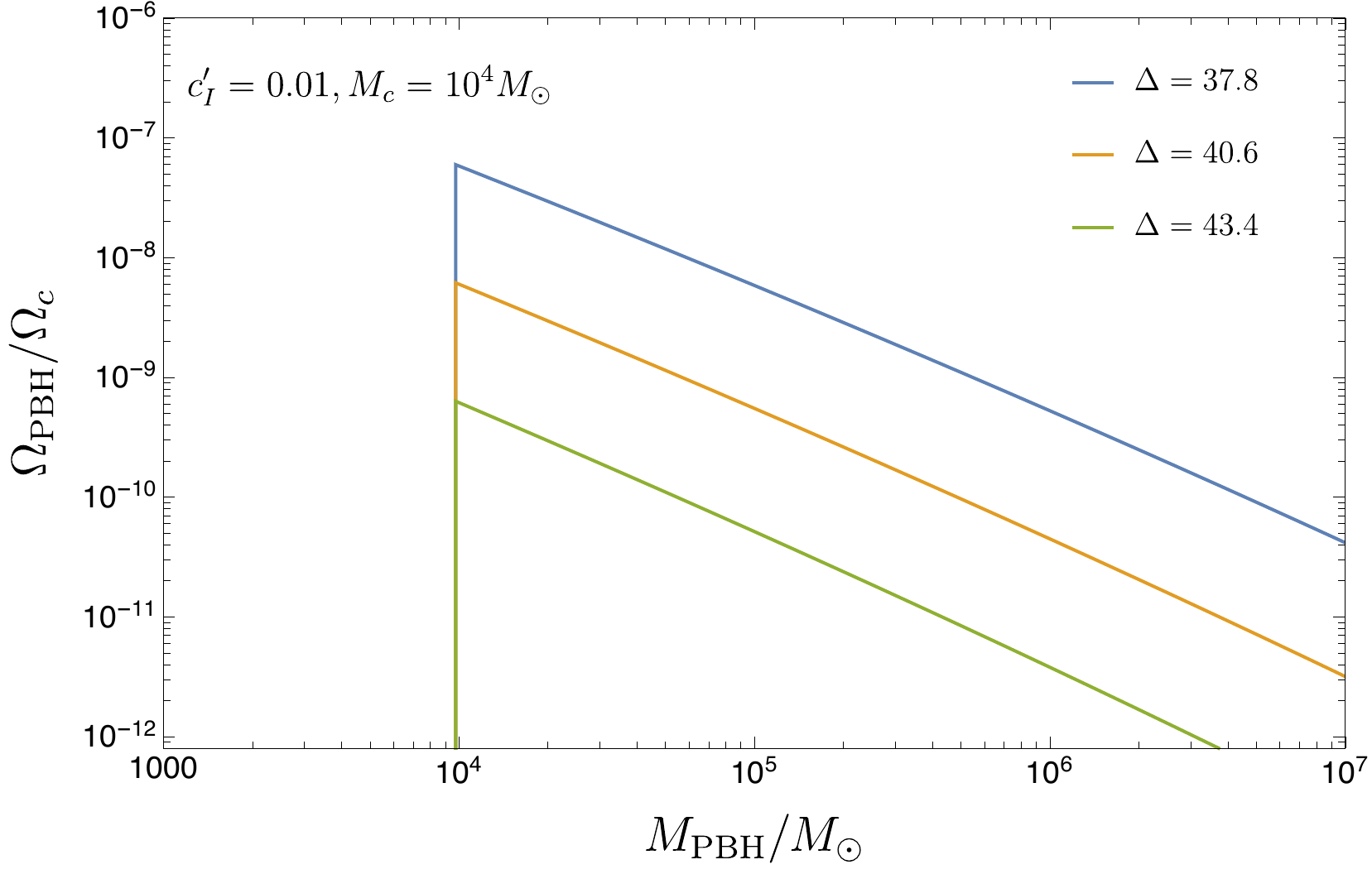}
    \includegraphics[clip,width=0.48\textwidth]{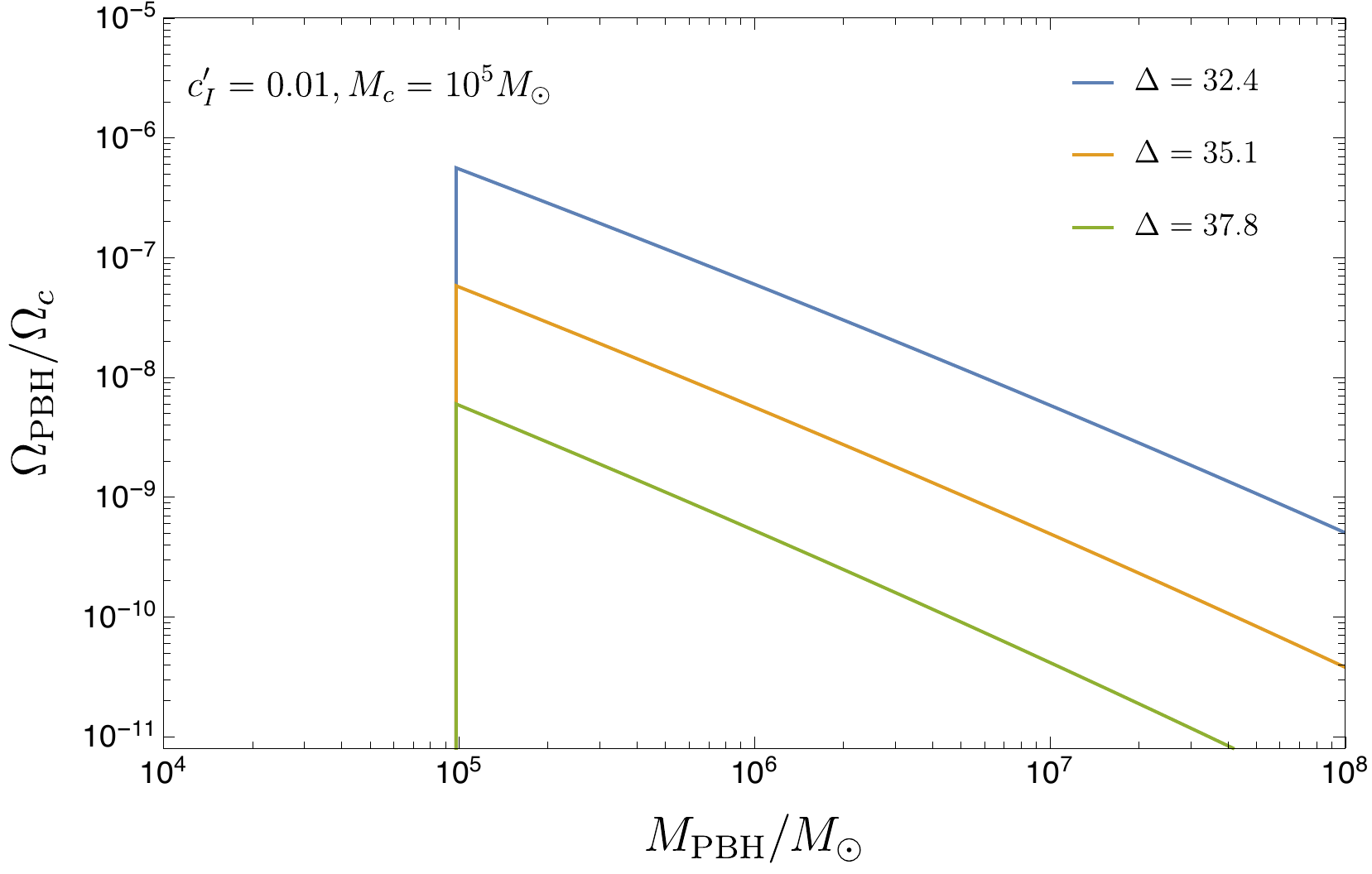}
    \caption{The PBH abundance.
            The left panel is the case with $M_c = 10^4M_{\odot}$ and the right panel is the case with $M_c = 10^5M_{\odot}$.
        In each panel, three lines correspond to $N_{\mathrm{gal}} = 0.1,0.01,0.001$~Mpc$^{-3}$ from top to bottom if $M_c \geq M_b$.}
    \label{fig:loglogQ}
  \end{center}
\end{figure}

Finally, we comment on the gauge-mediated SUSY breaking scenario.
As mentioned in the previous section, in this scenario, the coherent oscillation of the AD field in the HBBs can fragment into stable Q-balls.
Then, the density contrast between inside and outside the HBBs grows as $\propto T^{-1}$ and the HBBs which re-enter the horizon sufficiently later collapse into PBHs.
On the other hand, the residual Q-balls in the small HBBs which re-enter the horizon earlier and hence do not collapse into PBHs survive until now and they contribute to the current dark matter abundance.
Interestingly, for the appropriate parameters $c_I' \lesssim 0.01$, the PBHs originated from HBBs become the seed BHs of the SMBHs and at the same time the surviving Q-ball can account for the whole dark matter.
In Fig.~\ref{fig:loglogQ}, we show the abundance of PBHs with $c_I'=0.01$.
In the case of $M_c = 10^5M_{\odot}$ and $N{\mathrm{gal}}=0.1$~Mpc$^{-3}$ in Fig.~\ref{fig:loglogQ}, the surviving Q-balls account for almost all the dark matter.

\section{Conclusions}
\label{sec:conclusion}

In this paper, we have examined whether the HBBs produced in the modified AD mechanism proposed in Ref.~\cite{Hasegawa:2017jtk} account for the seed PBHs of the SMBHs in the galactic centers.
The modified AD mechanism in MSSM realizes inhomogeneous baryogenesis by taking into account the Hubble induced mass and the finite temperature effect, which leads to the formation of high baryon density regions called HBBs.
We have mainly studied the gravity-mediated SUSY breaking scenario.
In this scenario, after the QCD phase transition, the massive baryons inside the HBBs behave as non-relativistic matter and the HBBs have larger densities than the outside of the HBBs.
Because the density perturbations produced by the HBBs are highly non-Gaussian and HBBs are rare objects, the produced perturbations hardly constrained by the CMB $\mu$-distortion or the PTA experiments.

If the HBBs have sufficiently large densities when they re-enter the horizon, the HBBs gravitationally collapse into PBHs.
The density contrasts of the HBBs increase as the cosmic temperature deceases.
Therefore, only HBBs with large masses, which re-enter the horizon at late epochs, can form PBHs.
Thus, the condition for the PBH formation introduces a lower cut-off $M_c$ on the mass of the produced PBHs.
The mass distribution of the PBHs including the cut-off depends on the parameters of inflation and the potential for the AD field.
We have shown that the PBHs produced by this mechanism can have a reasonable number density and masses as the seed BHs of the SMBHs for appropriate values of the model parameters.

The small HBBs which do not collapse into PBHs contribute to the baryon asymmetry of the universe.
However, since the produced baryons are highly inhomogeneous, nucleosynthesis proceeds quite differently from the standard BBN.
Thus, the success of the BBN requires that the baryon asymmetry produced through HBBs should be much smaller than the observed one.
So, we need another mechanism or another AD field to produce the observed baryon asymmetry.

In the gauge-mediated SUSY breaking scenario, the stable Q-balls can contribute to both the PBHs and the dark matter abundance.
In our model, the PBHs produced from the Q-balls can have the reasonable number density and masses as the seed BHs of the SMBHs and, at the same time, the residual Q-balls can constitute the whole dark matter abundance.

Although the required mass distribution of the seed BHs of the SMBHs has some uncertainties associated with the number density of the galaxies and the mass growth of the SMBHs, the validity of these result are hardly affected because the mass distribution of the PBHs in our scenario can be easily controlled by varying the parameters.

\section*{Acknowledgments}
This work was supported by JSPS KAKENHI Grant Nos. 17H01131 (M.K.) and 17K05434 (M.K.),
MEXT KAKENHI Grant No. 15H05889 (M.K.), World Premier International Research Center
Initiative (WPI Initiative), MEXT, Japan (M.K., K.M.), and Program of Excellence in Photon Science (K.M.).

\bibliographystyle{JHEP}
\bibliography{SMBH}

\end{document}